# Many Sparse Cuts via Higher Eigenvalues

Anand Louis * Prasad Raghavendra * Prasad Tetali † Santosh Vempala *

**Abstract**

Cheeger's fundamental inequality states that any edge-weighted graph has a vertex subset $S$ such that its expansion (a.k.a. conductance) is bounded as follows:

$$\phi(S) \stackrel{\text{def}}{=} \frac{w(S, \bar{S})}{\min\{w(S), w(\bar{S})\}} \leqslant 2\sqrt{\lambda_2},$$

where $w$ is the total edge weight of a subset or a cut and $\lambda_2$ is the second smallest eigenvalue of the normalized Laplacian of the graph. Here we prove the following natural generalization: for any integer $k \in [n]$, there exist $ck$ disjoint subsets $S_1, \ldots, S_{ck}$, such that

$$\max_i \phi(S_i) \leqslant C\sqrt{\lambda_k \log k}$$

where $\lambda_i$ is the $i^{th}$ smallest eigenvalue of the normalized Laplacian and $c < 1, C > 0$ are suitable absolute constants. Our proof is via a polynomial-time algorithm to find such subsets, consisting of a spectral projection and a randomized rounding. As a consequence, we get the same upper bound for the small set expansion problem, namely for any $k$, there is a subset $S$ whose weight is at most a $\mathcal{O}(1/k)$ fraction of the total weight and $\phi(S) \leqslant C\sqrt{\lambda_k \log k}$. Both results are the best possible up to constant factors.

The underlying algorithmic problem, namely finding $k$ subsets such that the maximum expansion is minimized, besides extending sparse cuts to more than one subset, appears to be a natural clustering problem in its own right.

*College of Computing, Georgia Tech, Atlanta. Email : {anand.louis, raghavendra, vempala}@cc.gatech.edu
†School of Mathematics, Georgia Tech, Atlanta. Email : tetali@math.gatech.edu

# 1 Introduction

Given an edge-weighted graph $G = (V, E)$, a fundamental problem is to find a subset $S$ of vertices such that the total weight of edges leaving it is as small as possible compared to its size. This latter quantity, called expansion or conductance is defined as:

$$\phi_G(S) \stackrel{\text{def}}{=} \frac{w(S, \bar{S})}{\min\{w(S), w(\bar{S})\}}$$

where by $w(S)$ we denote the total weight of edges incident to vertices in $S$ and $w(S, T)$ is the total weight of edges between vertex subsets $S$ and $T$. The expansion of the graph $G$ is defined as

$$\phi_G \stackrel{\text{def}}{=} \min_{S: w(S) \leqslant 1/2} \phi_G(S).$$

Finding the optimal subset that minimizes expansion $\phi_G(S)$ is known as the sparsest cut problem. The expansion of a graph and the problem of approximating it (sparsest cut problem) have been highly influential in the study of algorithms and complexity, and have exhibited deep connections to many other areas of mathematics. In particular, motivated by its applications and the NP-hardness of the problem, the study of approximation algorithms for sparsest cut has been a very fruitful area of research.

In this line, the fundamental Cheeger's inequality (shown for graphs in [Alo86, AM85]) establishes a bound on expansion via the spectrum of the graph.

**Theorem 1.1** (Cheeger's Inequality ([Alo86, AM85]))**.** *For any graph $G$,*

$$\frac{\lambda_2}{2} \leqslant \phi_G \leqslant 2\sqrt{\lambda_2}$$

*where $\lambda_2$ is the second smallest eigenvalue of the normalized Laplacian* [1] *of $G$.*

The proof of Cheeger's inequality is algorithmic, using the eigenvector corresponding to the second smallest eigenvalue. This theorem and its many (minor) variants have played a major role in the design of algorithms as well as in understanding the limits of computation.

It has remained open to extend the sparsest cut problem to more than one subset and recover a suitable generalization of the above theorem. A natural way to do this is to consider a partition of the graph and some measure of the edges across parts. We survey such extensions in Section 1.1. Spectral partitioning algorithms are widely used in practice for their efficiency and the high quality of solutions that they often provide ([BS94, HL95]). An additional motivation for this work is from large unstructured data, where sparse cuts allow a large graph to be analyzed after applying the cuts. Here we focus on the following *$k$ sparse cuts* problem:

**$k$-sparse-cuts***: given an edge weighted graph $G = (V, E)$ and an integer $1 \leqslant k \leqslant n$, find $k$ disjoint subsets $S_1, \ldots, S_k$ of $V$ such that $\max_i \phi_G(S_i)$ is minimized.*

We note that sparsest cut corresponds to the case $k = 2$ as both sides of the cut are sparse. Notice that the sets $S_1, \ldots, S_k$ need not form a partition, i.e., there could be vertices that do not belong to any of the sets. The problem is a good way to model the existence of several well-formed *clusters* in a graph without requiring a partition, i.e., there could be vertices that do not participate in any cluster. A natural question is whether the problem is connected to higher eigenvalues of the graph. Infact it is not hard to show that the $k^{th}$ smallest eigenvalue of the normalized Laplacian of the graph gives a lower bound to the $k$-sparse cuts problem.

[1] See Section 1.2 for the definition of the normalized Laplacian of a graph.

**Proposition 1.2.** *For any edge-weighted graph $G = (V, E)$, for any integer $1 \leqslant k \leqslant |V|$, and for any $k$ disjoint subsets $S_1, \ldots, S_k \subset V$*

$$\max_i \phi_G(S_i) \geqslant \frac{\lambda_k}{2}$$

*where $\lambda_1, \ldots, \lambda_{|V|}$ are the eigenvalues of the normalized Laplacian of $G$.*

Our main result is an upper bound in terms of $\lambda_k$ analogous to Cheeger's inequality.

**Theorem 1.3.** *For any edge-weighted graph $G = (V, E)$, and any integer $1 \leqslant k \leqslant |V|$, there exist $ck$ disjoint subsets $S_1, \ldots, S_{ck}$ of vertices such that*

$$\max_i \phi_G(S_i) \leqslant C\sqrt{\lambda_k \log k}$$

*where $\lambda_1, \ldots, \lambda_{|V|}$ are the eigenvalues of the normalized Laplacian of $G$ and $c < 1, C$ are absolute constants. Moreover, these sets can be identified in polynomial time.*

The proof of this theorem is algorithmic and is based on spectral projection. Starting with the embedding given by the top $k$ eigenvectors of the (normalized) Laplacian of the graph, a simple randomized rounding procedure is used to produce $k$ vectors having disjoint support, and then a Cheeger cut is obtained from each of these vectors.

In general, one can not prove an upper bound better than $\mathcal{O}(\sqrt{\lambda_k \log k})$ for $k$-sparse cuts. This bound is matched by the family of *Gaussian graphs*. For a constant $\varepsilon \in (-1, 1)$, let $N_{k,\varepsilon}$ denote the infinite graph over $\mathbb{R}^k$ where the weight of an edge $(x, y)$ is the probability that two standard Gaussian random vectors $X, Y$ with correlation [2] $\varepsilon$ equal $x$ and $y$ respectively. The first $k$ eigenvalues of the Laplacian of $N_{k,\varepsilon}$ are at most $\varepsilon$ ([RST10b]). The following Lemma bounds the expansion of small sets in $N_{k,\varepsilon}$.

**Lemma 1.4** ([Bor85, RST10b])**.** *For any set $S \subset \mathbb{R}^k$ with Gaussian probability measure at most $1/k$,*

$$\phi_{N_{k,\varepsilon}}(S) \geqslant \Omega(\sqrt{\varepsilon \log k}).$$

Therefore, for any $k$ disjoint subsets $S_1, \ldots, S_k$ of $N_{k,\varepsilon}$ , $\max_i \phi_{N_{k,\varepsilon}}(S_i) \geqslant \Omega(\sqrt{\lambda_k \log k})$.

As an immediate consequence of Theorem 1.3, we get the following optimal bound on the small-set expansion problem ([RS10, RST10a]). This problem arose in the context of understanding the Unique Games Conjecture and has a close connection to it ([RS10, ABS10]).

**Corollary 1.5.** *For any edge-weighted graph $G = (V, E)$ and any integer $1 \leqslant k \leqslant |V|$, there is a subset $S$ with $w(S) = \mathcal{O}(1/k)w(V)$ and $\phi_G(S) \leqslant C\sqrt{\lambda_k \log k}$ for an absolute constant $C$.*

Theorem 1.3 also implies an upper bound of $\mathcal{O}(k\sqrt{\lambda_k \log k})$ on $\max_i \phi(S_i)$ for the case when the $ck$ sets are required to form a partition of the vertex set.

**Corollary 1.6.** *For any edge-weighted graph $G = (V, E)$ and any integer $1 \leqslant k \leqslant |V|$, there exists a partition of the vertex set $V$ into $ck$ parts $S_1, \ldots, S_{ck}$ such that*

$$\max_i \phi(S_i) \leqslant \mathcal{O}(k\sqrt{\lambda_k \log k})$$

*for an absolute constant $c$.*

Complementing the above bound, we show that for a $k$- partition $S_1, S_2, \ldots, S_k$, the quantity $\max_i \phi_G(S_i)$ cannot be bounded by $\mathcal{O}(\sqrt{\lambda_k}\mathsf{polylog}k)$ in general. We view this as further evidence suggesting that the $k$-sparse-cuts problem is the right generalization of sparsest cut.

**Theorem 1.7.** *There exists a family of graphs such that for any $k$-partition $(S_1, \ldots, S_k)$ of the vertex set*

$$\max_i \phi_G(S_i) \geqslant C \min\left\{\frac{k^2}{\sqrt{n}}, n^{\frac{1}{6}}\right\}\sqrt{\lambda_k}.$$

[2] $\varepsilon$ correlated Gaussians can be constructed as follows : $X \sim \mathcal{N}(0,1)^k$ and $Y \sim (1-\varepsilon)X + \sqrt{2\varepsilon - \varepsilon^2}Z$ where $Z \sim \mathcal{N}(0,1)^k$.

## 1.1 Related work

The classic sparsest cut problem has been extensively studied, and is known to be intimately connected to metric geometry [LLR95, AR98]. The lower and upper bounds on the sparsest cut given by Cheeger's inequality yields a $\mathcal{O}(\sqrt{\mathsf{OPT}})$ approximation algorithm for the sparsest cut problem. There are other known algorithms which have multiplicative approximation factors of $\mathcal{O}(\log n)$ via an LP relaxation [LR99, AR98] and $\mathcal{O}(\sqrt{\log n})$ via a semi-definite relaxation [ARV04]). In many contexts, and in practice, the eigenvector approach is often preferred in spite of a higher worst-case approximation factor.

The small set expansion problem – a generalization of sparsest cut has received much attention owing to its connection to the unique games conjecture. Here the goal is to find a subset of minimum sparsity that contains at most a $\mathcal{O}(1/k)$ fraction of the vertices. In terms of eigenvalues, this quantity was shown to be upper bounded by $\mathcal{O}(\sqrt{\lambda_{k^2} \log k})$ in [LRTV11], and by $\mathcal{O}(\sqrt{\lambda_{k^{100}} \log_k n})$ in [ABS10]. Using a semidefinite programming relaxation, [RST10a] obtain an algorithm that outputs a small set with expansion at most $\sqrt{\mathsf{OPT} \log k}$ where $\mathsf{OPT}$ is the sparsity of the optimal set of size at most $\mathcal{O}(1/k)$. Bansal et.al. [BFK$^+$11] obtained an $\mathcal{O}(\sqrt{\log n \log k})$ approximation algorithm for the small set expansion problem using a semidefinite programming relaxation.

A second generalization that has been studied is that of partitioning the vertex set of a graph into $k$ parts so as to minimize the sparsity of the partition (defined as the ratio of the weight of edges between parts to the total weight of edges incident to the smallest $k-1$ parts). [LRTV11] give a polynomial time algorithm for finding a $k$-partition with sparsity at most $8\sqrt{\lambda_k} \log k$. Their proof is based on a simple recursive algorithm. Closer to this is the $(\alpha, \varepsilon)$-clustering problem that asks for a partition where each part has conductance at least $\alpha$ and the total weight of edges removed is minimized. [KVV04] give a recursive algorithm to obtain a bi-criteria approximation to the $(\alpha, \varepsilon)$-clustering problem. Indeed recursive algorithms are one of most commonly used techniques in practice for graph multi-partitioning. However, we show that partitioning a graph into $k$ pieces using a simple recursive algorithm can yield as many $k(1 - o(1))$ sets with expansion much larger than $\sqrt{\lambda_k}\mathsf{polylog} k$ (See Appendix A).

[GT11] studied a close variant of the problem we consider, and show that every graph $G$ has a $k$ partition such that each part has expansion at most $\mathcal{O}(k^6\sqrt{\lambda_k})$. Other generalizations of the sparsest cut problem have been considered for special classes of graphs ([BLR10, Kel06, ST96]).

A randomized rounding step similar to the one in our algorithm was used previously in the context of rounding semidefinite programs for unique games ([CMM06]).

## 1.2 Notation

For a graph $G = (V, E)$, we let $A$ be its (weighted) adjacency matrix and $d_i$ be the (weighted) degree of vertex $i$. We use $D$ to denote the diagonal matrix with $D_{ii} = d_i$. The *normalized* Laplacian of a graph defined as

$$\mathcal{L}_G \stackrel{\text{def}}{=} D^{-\frac{1}{2}}(D - A)D^{-\frac{1}{2}}$$

We let $0 = \lambda_1 \leqslant \lambda_2 \leqslant \ldots \lambda_n$ denote the eigenvalues of $\mathcal{L}_G$ and $v'_1, v'_2, \ldots, v'_n$ denote the corresponding eigenvectors. Let $v_i \stackrel{\text{def}}{=} D^{-\frac{1}{2}} v'_i$ for each $i \in [n]$. Therefore, $v'^T_i \mathcal{L}_G v'_i = \sum_{u \sim w} (v_i(u) - v_j(w))^2$. Since $\forall i \neq j$ $\langle v'_i, v'_j \rangle = 0$, $\sum_l d_l v_i(l) v_j(l) = 0$

## 1.3 Organization

We begin with our main algorithm used in the proof of Theorem 1.3 in Section 2. In Section 3.1 we give an overview of the proof of Theorem 1.3. We present some preliminaries needed for our proof in Section 3.2 and in Section 3.3 we present the proof of Theorem 1.3. In Section 4 we present the proof of Proposition 1.2, and finally in Section 5 we present the proof of Theorem 1.7.

## 2 Algorithm

Our algorithm for finding $\Theta(k)$ sparse cuts appears in Figure 1.

Input : Graph $G = (V, E)$, parameter $k$.

1. **Spectral projection.** Let $V = [v_1, \ldots, v_k]$ be an $n \times k$ matrix whose columns are the top $k$ eigenvectors of the normalized Laplacian $\mathcal{L}_G$; let $u_1, \ldots, u_n$ be the rows of $V$.
2. **Randomized rounding.** Pick $k$ independent Gaussian vectors $g_1, g_2, \ldots, g_k \sim \mathcal{N}(0, 1)^k$. Construct vectors $h_1, h_2, \ldots, h_k \in \mathbb{R}^n$ as follows:
$$h_i(j) = \begin{cases} \langle u_j, g_i \rangle \text{ if } i = \mathsf{argmax}_{i \in [k]} \{\langle u, g_i \rangle\} \\ 0 \text{ otherwise.} \end{cases}$$
3. **Cheeger cuts.** For $j = 1, \ldots, k$, sort the cooordinates of $h_j$ according to their magnitude, and pick the subset of lowest sparsity in this linear ordering.
4. Output all subsets of expansion smaller than $\mathcal{O}(\sqrt{\lambda_k \log k})$.

Figure 1: The Many-sparse-cuts Algorithm

## 3 Analysis

### 3.1 Outline

Notice that the vectors $h_1, h_2, \ldots, h_k$ have disjoint support since for each coordinate $j$, exactly one of the $\langle u_j, g_i \rangle$ is maximum. Therefore, the Cheeger cuts obtained by the vectors $h_i$ yield $k$ disjoint sets. It is sufficient to show that a constant fraction of the sets so produced have small expansion.

As first attempt to proving the upper bound in Theorem 1.3, one could first try to bound the Rayleigh quotient of the vectors $\{h_i\}$ by $\mathcal{O}(\lambda_k \log k)$ (say for a constant fraction of vectors $h_i$). This would imply that the corresponding sets would have value $\mathcal{O}(\sqrt{\lambda_k \log k})$ by following the proof of Cheeger's inequality. Unfortunately, we note that the Rayleigh quotient of the vectors obtained could themselves be as high as $\Omega(\sqrt{\lambda_k \log k})$, and using the proof of Cheeger's inequality this would at best yield a bound of $\mathcal{O}((\lambda_k \log k)^{\frac{1}{4}})$ on the expansion of the sets obtained. Therefore, in our proof we directly analyze the quality of the Cheeger cuts finally output by the algorithm.

We will show that for each $i \in \{1, \ldots, k\}$, the vector $h_i$ has a constant probability of yielding a cut with small expansion. The outline of the proof is as follows. Let $f$ denote the vector $h_1$. The choice of the index 1 is arbitrary and the same analysis is applicable to all other indices $i \in [m]$.

The quality of the Cheeger cut obtained from $f$ can be upper bounded using the following standard Lemma used in most known algorithms for sparsest cut. A proof of this Lemma can be found in [Chu97].

**Lemma 3.1.** *Let $x \in \mathbb{R}^n$ be a vector such that $\frac{\sum_{i \sim j} |x_i - x_j|}{\sum_i d_i x_i} \leqslant \delta$. Then one of the level sets, say $S$, of the vector $x$ has $\phi_G(S) \leqslant 2\delta$.*

Applying Lemma 3.1, the expansion of the set retrieved from $f = h_1$ is upper bounded by,

$$\frac{\sum_{(i,j) \in E} |f_i^2 - f_j^2|}{\sum_i d_i f_i^2}.$$

Both the numerator and denominator are random variables depending on the choice of random Gaussians $g_1, \ldots, g_k$. It is a fairly straightforward calculation to bound the expected value of the denominator.

**Lemma 3.2.**

$$2 \log k \leqslant \mathbb{E}\left[\sum_i d_i f_i^2\right] \leqslant 4 \log k.$$

Bounding the expected value of the numerator is more subtle. We show the following bound on the expected value of the numerator.

**Lemma 3.3.**

$$\mathbb{E}\left[\sum_{i \sim j} |f_i^2 - f_j^2|\right] \leqslant 8(2c_1 + 1)(\sqrt{\lambda_k} \log^{\frac{3}{2}} k).$$

Notice that the ratio of their expected values is $\mathcal{O}(\sqrt{\lambda_k \log k})$, as intended. To control the ratio of the two quantities, the numerator is to be bounded from above, and the denominator is to be bounded from below. A simple Markov inequality can be used to upper bound the probability that the numerator is much larger than its expectation. To control the denominator, we bound its variance. Specifically, we will show the following bound on the variance of the denominator.

**Lemma 3.4.**

$$\mathsf{Var}\left[\sum_i d_i f_i^2\right] \leqslant 32 \log^2 k.$$

The above moment bounds are sufficient to conclude that with constant probability, the ratio $\frac{\sum_{(i,j) \in E} |f_i^2 - f_j^2|}{\sum_i d_i f_i^2}$ is within a constant factor of $\mathcal{O}(\sqrt{\lambda_k \log k})$. Therefore, with constant probability over the choice of the Gaussians $g_1, \ldots, g_k$, $\Omega(k)$ of the vectors $h_1, \ldots h_k$ yield sets of expansion $\mathcal{O}(\sqrt{\lambda_k \log k})$.

## 3.2 Preliminaries

We collect here some useful properties of the spectral embedding, the randomized rounding, and a well-known analysis of Cheeger cuts.

### Spectral Embedding

**Lemma 3.5.** *(Spectral embedding)*

1.
$$\frac{\sum_{i \sim j} \|u_i - u_j\|^2}{\sum_i d_i \|u_i\|^2} \leqslant \lambda_k$$

2.
$$\sum_i d_i \|u_i\|^2 = k$$

3.
$$\sum_{i,j} d_i d_j \langle u_i, u_j \rangle^2 = k.$$

*Proof of (3).*

$$\begin{aligned}\sum_{i,j} d_i d_j \langle u_i, u_j\rangle^2 &= \sum_{i,j} d_i d_j \left(\sum_{t=1}^{k} u_i(t)u_j(t)\right)^2 \\ &= \sum_{i,j} d_i d_j \sum_{t_1,t_2} u_i(t_1)u_j(t_1)u_i(t_2)u_j(t_2) \\ &= \sum_{t_1,t_2}\sum_{i,j} d_i d_j u_i(t_1)u_j(t_1)u_i(t_2)u_j(t_2) \\ &= \sum_{t_1,t_2}\left(\sum_i d_i u_i(t_1)u_i(t_2)\right)^2\end{aligned}$$

Since $\sqrt{d_i}u_i(t_1)$ is the entry to corresponding to vertex $i$ in the $t_1^{th}$ eigenvector, $\sum_i d_i u_i(t_1)u_i(t_2)$ is equal to the inner product of the $t_1^{th}$ and $t_2^{th}$ eigenvectors of $\mathcal{L}$, which is equal to 1 only when $t_1 = t_2$ and is equal to 0 otherwise. Therefore,

$$\sum_{i,j} d_i d_j \langle u_i, u_j\rangle^2 = \sum_{t_1,t_2} \mathbb{I}\left[t_1 = t_2\right] = k$$

□

Next, we recall the one-sided Chebychev inequality.

**Fact 3.6** (One-sided Chebychev Inequality)**.** *For a random variable $X$ with mean $\mu$ and variance $\sigma^2$ and any $t > 0$,*

$$\mathbb{P}\left[X < \mu - t\sigma\right] \leqslant \frac{1}{1+t^2}.$$

**Properties of Gaussian Variables** The next few facts are folklore about Gaussians. Let $t_{1/k}$ denote the $(1/k)^{th}$ cap of a standard normal variable, i.e., $t_{1/k} \in \mathbb{R}$ is the number such that for a standard normal random variable $X$, $\mathbb{P}\left[X \geqslant t_{1/k}\right] = 1/k$.

**Fact 3.7.** *For a standard normal random variable $X$ and for every $t > 0$,*

$$t_{1/k} \approx \sqrt{2\log k - \log\log k}$$

**Fact 3.8.** *Let $X_1, X_2, \ldots, X_k$ be $k$ independent standard normal random variables. Let $Y$ be the random variable defined as $Y \stackrel{\text{def}}{=} \max\{X_i | i \in [k]\}$. Then*

1. $t_{1/k} \leqslant \mathbb{E}\left[Y\right] \leqslant 2\sqrt{\log k}$
2. $\mathbb{E}\left[Y^2\right] \leqslant 4\log k$
3. $\mathbb{E}\left[Y^4\right] \leqslant 4e\log^2 k$
4. *For any constant $\varepsilon$,* $\mathbb{P}\left[Y \leqslant (1-\varepsilon)t_{1/k}\right] \leqslant \frac{1}{e^{k^{2\varepsilon}}}$

*Proof.* For any $Z_1, \ldots, Z_k \in \mathbb{R}^+$ and any $p \in \mathbb{Z}^+$, we have $\max_i Z_i \leqslant (\sum_i Z_i^p)^{\frac{1}{p}}$. Now $Y^4 = (\max_i X_i)^4 \leqslant \max_i X_i^4$.

$$\begin{aligned}\mathbb{E}\left[Y^4\right] &\leqslant \mathbb{E}\left[\left(\sum_i X_i^{4p}\right)^{\frac{1}{p}}\right] \leqslant \left(\mathbb{E}\left[\left(\sum_i X_i^{4p}\right)\right]\right)^{\frac{1}{p}} \quad \text{( Jensen's Inequality )} \\ &\leqslant \left(\sum_i (\mathbb{E}\left[X_i^2\right])\frac{(4p)!}{(2p)!2^{2p}}\right)^{\frac{1}{p}} \leqslant 4p^2 k^{\frac{1}{p}} \quad \text{(using } (4p)!/(2p)! \leqslant (4p)^{2p} \text{ )}\end{aligned}$$

Picking $p = \log k$ gives $\mathbb{E}\left[Y^4\right] \leqslant 4e\log^2 k$.

Therefore $\mathbb{E}\left[Y^2\right] \leqslant \sqrt{\mathbb{E}\left[Y^4\right]} \leqslant 4\log k$ and $\mathbb{E}\left[Y\right] \leqslant \sqrt{\mathbb{E}\left[Y^2\right]} \leqslant 2\sqrt{\log k}$.

And,

$$\mathbb{P}\left[Y \leqslant (1-\varepsilon)t_{1/k}\right] \leqslant \left(1 - \frac{1}{k^{1-2\varepsilon}}\right)^k \leqslant \frac{1}{e^{k^{2\varepsilon}}}$$

□

**Fact 3.9.** *Let $X_1, \ldots, X_k$ and $Y_1, \ldots, Y_k$ be i.i.d. standard normal random variables such that for all $i \in [k]$, the covariance of $X_i$ and $Y_i$ is $1 - \varepsilon^2$. Then*

$$\mathbb{P}\left[\mathsf{argmax}_i X_i \neq \mathsf{argmax}_i Y_i\right] \leqslant c_1 \left(\varepsilon\sqrt{\log k}\right)$$

*for some absolute constant $c_1$.*

We refer the reader to [CMM06] for the proof of a more general claim.

## 3.3 Main Proof

Let $f$ denote the vector $h_1$. The choice of the index 1 is arbitrary and the same analysis is applicable to all other indices $i \in [m]$. We first separately bound the expectations of the numerator and denominator of the sparsity of each cut, and then the variance of the denominator. The proofs of these bounds will follow their application in the proof of our main theorem.

**Expectation of the Denominator** Bounding the expectation of the denominator is a straightforward calculation as shown below.

**Lemma 3.10** (Restatement of Lemma 3.2)**.**

$$2\log k \leqslant \mathbb{E}\left[\sum_i d_i f_i^2\right] \leqslant 4\log k.$$

*Proof of Lemma 3.2.* For any $i \in [n]$, recall that

$$f_i = \begin{cases} \|u_i\|\langle \tilde{u_i}, g_1\rangle \text{ if } \langle \tilde{u_i}, g_1\rangle \geqslant \langle \tilde{u_i}, g_j\rangle \ \forall j \in [k] \\ 0 \text{ otherwise.} \end{cases}$$

The first case happens with probability $1/k$ and so $f_i = 0$ with the remaining probability. Therefore, using Fact 3.8,

$$2\|u_i\|^2 \log k/k \leqslant \mathbb{E}\left[f_i^2\right] \leqslant 4\|u_i\|^2 \log k/k$$

and hence

$$2\log k \leqslant \mathbb{E}\left[\sum_i d_i f_i^2\right] \leqslant 4\log k$$

using $\sum_i d_i \|u_i\|^2 = k$ from Lemma 3.5. □

**Expectation of the Numerator** For bounding the expecation of the numerator we will need some preparation. We will make use of the following proposition which relates distance between two vectors to the distance between the unit vectors in the corresponding directions.

**Proposition 3.11.** *For any two non zero vectors $u_i$ and $u_j$, if $\tilde{u_i} = u_i/\|u_i\|$ and $\tilde{u_j} = u_j/\|u_j\|$ then*

$$\|\tilde{u_i} - \tilde{u_j}\|\sqrt{\|u_i\|^2 + \|u_j\|^2} \leqslant 2\|u_i - u_j\|$$

*Proof.* Note that $2\|u_i\|\|u_j\| \leqslant \|u_i\|^2 + \|u_j\|^2$. Hence,

$$\begin{aligned} \|\tilde{u_i} - \tilde{u_j}\|^2(\|u_i\|^2 + \|u_j\|^2) &= (2 - 2\langle \tilde{u_i}, \tilde{u_j}\rangle)(\|u_i\|^2 + \|u_j\|^2) \\ &\leqslant 2(\|u_i\|^2 + \|u_j\|^2 - (\|u_i\|^2 + \|u_j\|^2)\langle \tilde{u_i}, \tilde{u_j}\rangle) \end{aligned}$$

If $\langle \tilde{u_i}, \tilde{u_j}\rangle \geqslant 0$, then

$$\|\tilde{u_i} - \tilde{u_j}\|^2(\|u_i\|^2 + \|u_j\|^2) \leqslant 2(\|u_i\|^2 + \|u_j\|^2 - 2\|u_i\|\|u_j\|\langle \tilde{u_i}, \tilde{u_j}\rangle) \leqslant 2\|u_i - u_j\|^2$$

Else if $\langle \tilde{u_i}, \tilde{u_j}\rangle < 0$, then

$$\|\tilde{u_i} - \tilde{u_j}\|^2(\|u_i\|^2 + \|u_j\|^2) \leqslant 4(\|u_i\|^2 + \|u_j\|^2 - 2\|u_i\|\|u_j\|\langle \tilde{u_i}, \tilde{u_j}\rangle) \leqslant 4\|u_i - u_j\|^2$$

□

We will also make use of the following propositions which bounds the expected value of a conditioned random variable.

**Proposition 3.12.** *For indices $i \neq j$*

$$\mathbb{E}\left[\langle u_i, g_1\rangle^2 | f_j > 0\right] \mathbb{P}\left[f_j > 0\right] \leqslant \frac{4}{k}\|u_i\|^2 \log k$$

*Proof.*

$$\begin{aligned} \mathbb{E}\left[\langle u_i, g_1\rangle^2 | f_j > 0\right] \mathbb{P}\left[f_j > 0\right] &\leqslant \mathbb{E}\left[\max_{p\in[k]}\langle u_i, g_p\rangle^2 | \langle u_j, g_1\rangle \geqslant \langle u_j, g_l\rangle \forall l \in [k]\right] \\ &\qquad \mathbb{P}\left[\langle u_j, g_1\rangle \geqslant \langle u_j, g_l\rangle \forall l \in [k]\right] \\ &= \frac{1}{k}\sum_{q\in[k]} \mathbb{E}\left[\max_{p\in[k]}\langle u_i, g_p\rangle^2 | \langle u_j, g_q\rangle \geqslant \langle u_j, g_l\rangle \forall l \in [k]\right] \\ &\qquad \mathbb{P}\left[\langle u_j, g_q\rangle \geqslant \langle u_j, g_l\rangle \forall l \in [k]\right] \\ &= \frac{1}{k}\mathbb{E}\left[\max_{p\in[k]}\langle u_i, g_p\rangle^2\right] \\ &= \frac{4}{k}\|u_i^2\| \log k \end{aligned}$$

□

Similiarly, we also prove the following proposition.

**Proposition 3.13.** *For indices $i \neq j$*

$$\mathbb{E}\left[\langle u_i, g_1\rangle^2 | f_i > 0 \text{ and } f_j = 0\right] \mathbb{P}\left[f_j = 0\right] \leqslant 4\left(1 - \frac{1}{k}\right)\|u_i\|^2 \log k$$

We will need another Lemma that is a direct consequence of Fact 3.9 about the maximum of $k$ i.i.d normal random variables.

**Proposition 3.14.** *For any $i, j \in [n]$,*

$$\mathbb{P}\left[f_i > 0 \text{ and } f_j = 0\right] \leqslant c_1\left(\|\tilde{u_i} - \tilde{u_j}\|\frac{\sqrt{\log k}}{k}\right).$$

We are now ready to bound the expectation of the numerator, we restate the Lemma for the convenience of the reader.

**Lemma 3.15.** *(Restatement of Lemma 3.3)*

$$\mathbb{E}\left[\sum_{i\sim j}|f_i^2 - f_j^2|\right] \leqslant 8(2c_1+1)(\sqrt{\lambda_k}\log^{\frac{3}{2}} k).$$

*Proof of Lemma 3.3.* From Fact 3.8, $\mathbb{E}\left[f_i^2 | f_i > 0\right] \leqslant 4\|u_i\|^2 \log k$. Therefore,

$$\begin{aligned}\mathbb{E}\left[|f_i^2 - f_j^2||f_i, f_j > 0\right] &= \mathbb{E}\left[|\langle u_i, g_1\rangle^2 - \langle u_j, g_1\rangle^2||f_i, f_j > 0\right] \\ &= \mathbb{E}\left[|\langle u_i - u_j, g_1\rangle\langle u_i + u_j, g_1\rangle||f_i, f_j > 0\right] \\ &\leqslant \sqrt{\mathbb{E}\left[\langle u_i - u_j, g_1\rangle^2 | f_i, f_j > 0\right]}\sqrt{\mathbb{E}\left[\langle u_i + u_j, g_1\rangle^2 | f_i, f_j > 0\right]}\end{aligned}$$

Now,

$$\mathbb{E}\left[\langle u_i - u_j, g_1\rangle^2 | f_i, f_j > 0\right]\mathbb{P}\left[f_i, f_j > 0\right] = \int_{f_i, f_j > 0} \langle u_i - u_j, g_1\rangle^2$$

$$\leqslant \int_{f_i > 0} \langle u_i - u_j, g_1\rangle^2 = \mathbb{E}\left[\langle u_i - u_j, g_1\rangle^2 | f_i > 0\right]\mathbb{P}\left[f_i > 0\right]$$

$$= \frac{1}{k}\sum_{p\in[k]}\mathbb{E}\left[\max_l \langle u_i - u_j, g_l\rangle^2 | \langle u_i, g_p\rangle \geqslant \langle u_i, g_l\rangle \forall l \in [k]\right]\mathbb{P}\left[\langle u_i, g_p\rangle \geqslant \langle u_i, g_l\rangle \forall l \in [k]\right]$$

$$= \frac{1}{k}\mathbb{E}\left[\max_l \langle u_i - u_j, g_l\rangle^2\right] \leqslant \frac{4}{k}\|u_i - u_j\|^2 \log k$$

Similiarly, we get

$$\mathbb{E}\left[\langle u_i + u_j, g_1\rangle^2 | f_i, f_j > 0\right]\mathbb{P}\left[f_i, f_j > 0\right] \leqslant \frac{4}{k}\|u_i + u_j\|^2 \log k$$

Therefore, we get

$$\mathbb{E}\left[|f_i^2 - f_j^2||f_i, f_j > 0\right]\mathbb{P}\left[f_i, f_j > 0\right] \leqslant \frac{4}{k}\|u_i - u_j\|\|u_i + u_j\| \log k$$

From Proposition 3.14,

$$\mathbb{P}\left[f_i > 0 \text{ and } f_j = 0\right] = \mathbb{P}\left[f_j > 0 \text{ and } f_i = 0\right] \leqslant c_1(\|\tilde{u_i} - \tilde{u_j}\|\sqrt{\log k}/k).$$

Therefore,

$$\begin{aligned}\mathbb{E}\left[f_i^2 - f_j^2 | f_i > 0, f_j = 0\right]\mathbb{P}\left[f_i > 0, f_j = 0\right] &= \mathbb{E}\left[\langle u_1, g\rangle^2 | f_i > 0, f_j = 0\right]\mathbb{P}\left[f_j = 0\right]\frac{\mathbb{P}\left[f_i > 0, f_j = 0\right]}{\mathbb{P}\left[f_j = 0\right]} \\ &\leqslant 4(1 - \frac{1}{k})\|u_i\|^2 \log k \frac{\mathbb{P}\left[f_i > 0, f_j = 0\right]}{1 - 1/k} \quad (\text{ Proposition 3.13 }) \\ &= \frac{4c_1}{k}\log^{\frac{3}{2}} k \|u_i\|^2 \|\tilde{u_i} - \tilde{u_j}\|\end{aligned}$$

Similarly,

$$\mathbb{E}\left[f_j^2 - f_i^2 | f_j > 0, f_i = 0\right]\mathbb{P}\left[f_j > 0, f_i = 0\right] \leqslant \frac{4c_1}{k}\log^{\frac{3}{2}} k \|u_j\|^2 \|\tilde{u_i} - \tilde{u_j}\|\sqrt{\log k}.$$

Next,

$$
\begin{aligned}
\mathbb{E}\left[\sum_{i\sim j}|f_i^2-f_j^2|\right] &\leqslant \frac{4\log k}{k}\left(\sum_{i\sim j}\|u_i-u_j\|\|u_i+u_j\|+c_1(\|u_i\|^2+\|u_j\|^2)\|\tilde{u_i}-\tilde{u_j}\|\sqrt{\log k}\right)\\
&\leqslant \frac{4\log k}{k}\left(\sum_{i\sim j}\|u_i-u_j\|(\|u_i\|+\|u_j\|)+2c_1\|u_i-u_j\|\sqrt{\|u_i\|^2+\|u_j\|^2}\sqrt{\log k}\right)\\
&\quad (\text{ from Lemma 3.11})\\
&\leqslant \frac{4(2c_1+1)\log^{\frac{3}{2}}k}{k}\sum_{i\sim j}\|u_i-u_j\|(\|u_i\|+\|u_j\|)\\
&\leqslant \frac{4(2c_1+1)\log^{\frac{3}{2}}k}{k}\sqrt{\sum_{i\sim j}(\|u_i-u_j\|)^2}\sqrt{\sum_{i\sim j}(\|u_i\|+\|u_j\|)^2} \quad \text{(Cauchy-Schwartz inequality)}\\
&\leqslant \frac{4(2c_1+1)\log^{\frac{3}{2}}k}{k}\sqrt{\lambda_k\sum_i d_i\|u_i\|^2}\sqrt{4\sum_i d_i\|u_i\|^2}\\
&= \frac{8(2c_1+1)\log^{\frac{3}{2}}k}{k}\sqrt{\lambda_k}\sum_i d_i\|u_i\|^2\\
&= 8(2c_1+1)\log^{\frac{3}{2}}k\sqrt{\lambda_k}
\end{aligned}
$$

□

**Variance of the Denominator** Here too we will need some groundwork.

Let $\mathcal{G}$ denote the Gaussian space. The Hermite polynomials $\{H_i\}_{i\in\mathbb{Z}_{\geqslant 0}}$ form an orthonormal basis for real valued functions over the Gaussian space $\mathcal{G}$, i.e., $\mathbb{E}_{g\in\mathcal{G}}[H_i(g)H_j(g)] = 1$ if $i = j$ and 0 otherwise. The $k$-wise tensor product of the Hermite basis forms an orthonormal basis for functions over $\mathcal{G}^k$. Specifically, for each $\alpha \in \mathbb{Z}^k_{\geqslant 0}$ define the polynomial $H_\alpha$ as

$$H_\alpha(x_1,\ldots,x_k) = \prod_{i=1}^{k} H_{\alpha_i}(x_i).$$

The functions $\{H_\alpha\}_{\alpha\in\mathbb{Z}^k_{\geqslant 0}}$ form an orthonormal basis for functions over $\mathcal{G}^k$. The degree of the polynomial $H_\alpha(x)$ denote by $|\alpha|$ is $|\alpha| = \sum_i \alpha_i$.

The Hermite polynomials form a complete eigenbasis for the noise operator on the Gaussian space (Ornstein-Uhlenbeck operator). In particular, they are known to satisfy the following property (see e.g. the book of Ledoux and Talagrand [LT91], Section 3.2).

**Fact 3.16.** *Let $(g_i,h_i)_{i=1}^k$ be $k$ independent samples from two $\rho$-correlated Gaussians, i.e., $\mathbb{E}[g_i^2] = \mathbb{E}[h_i^2] = 1$, and $\mathbb{E}[g_ih_i] = \rho$. Then for all $\alpha \in \mathbb{Z}^k_{\geqslant 0}$,*

$$\mathbb{E}[H_\alpha(g_1,\ldots,g_k)H_{\alpha'}(h_1,\ldots,h_k)] = \rho^{|\alpha|} \textit{ if } \alpha = \alpha' \textit{ and } 0 \textit{ otherwise}$$

Let $A : \mathcal{G}^k \longrightarrow \mathbb{R}$ be the function defined as follows,

$$A(x) = \begin{cases} x_1^2 & \text{if } (x_1 \geqslant x_i \forall i \in [k]) \vee (x_1 \leqslant x_i \forall i \in [k]) \\ 0 & \text{otherwise} \end{cases}$$

We know that

$$\mathbb{E}[A] \leqslant \frac{4\log k}{k} \text{ and } \mathbb{E}[A^2] \leqslant \frac{16\log^2 k}{k} \text{ (Fact 3.8).}$$

**Lemma 3.17.** *Let $u, v$ be unit vectors and $g_1, \ldots, g_k$ be i.i.d Gaussian vectors. Then,*

$$\mathbb{E}[A(\langle u, g_1\rangle, \ldots, \langle u, g_k\rangle)A(\langle v, g_1\rangle, \ldots, \langle v, g_k\rangle)] \leqslant \frac{16\log^2 k}{k}\left(\langle u, v\rangle^2 + \frac{1}{k}\right)$$

*Proof.* The function $A$ on the Gaussian space can be written in the Hermite expansion $A(x) = \sum_\alpha A_\alpha H_\alpha(x)$ such that

$$\sum_\alpha A_\alpha^2 = \mathbb{E}[A^2] \leqslant \frac{16\log^2 k}{k}.$$

Using Fact 3.16, we can write

$$\mathbb{E}[A(\langle u, g_1\rangle, \ldots, \langle u, g_k\rangle)A(\langle v, g_1\rangle, \ldots, \langle v, g_k\rangle)] = (\mathbb{E}[A])^2 + \sum_{\alpha \in \mathbb{Z}^k_{\geqslant 0}, |\alpha|>0} A_\alpha^2 \rho^{|\alpha|}$$

where $\rho = \langle u, v\rangle$. Since $A$ is an even function, only the even degree coefficients are non-zero, i.e., $A_\alpha = 0$ for all $|\alpha|$ odd. Along with $\rho \leqslant 1$, this implies that

$$\begin{aligned}\mathbb{E}[A(\langle u, g_1\rangle, \ldots, \langle u, g_k\rangle)A(\langle v, g_1\rangle, \ldots, \langle v, g_k\rangle)] &\leqslant (\mathbb{E}[A])^2 + \rho^2\left(\sum_{\alpha, |\alpha|\geqslant 2} A_\alpha^2\right) \qquad \text{where } \rho = \langle u, v\rangle \\ &\leqslant \frac{16\log^2 k}{k^2} + \langle u, v\rangle^2 \frac{16\log^2 k}{k}\end{aligned}$$

□

Now, we will show a bound on the variance of the denominator.

*Proof of Lemma 3.4.*

$$\begin{aligned}\mathbb{E}\left[\sum_{i,j} d_i d_j f_i^2 f_j^2\right] &= \sum_{i,j} d_i d_j \|u_i\|^2 \|u_j\|^2\, \mathbb{E}\left[\frac{f_i^2}{\|u_i\|^2}\frac{f_j^2}{\|u_j\|^2}\right] \\ &\leqslant \sum_{i,j} d_i d_j \|u_i\|^2 \|u_j\|^2\, \mathbb{E}\left[A(\langle \tilde{u}_i, g_1\rangle, \ldots, \langle \tilde{u}_i, g_k\rangle)A(\langle \tilde{u}_j, g_1\rangle, \ldots, \langle \tilde{u}_j, g_k\rangle))\right] \\ &\leqslant \sum_{i,j} d_i d_j \|u_i\|^2 \|u_j\|^2 \cdot \left(\frac{16\log^2 k}{k}(\langle \tilde{u}_i, \tilde{u}_j\rangle^2 + \frac{1}{k})\right) \qquad \text{(Lemma 3.17)} \\ &\leqslant \frac{16\log^2 k}{k} \cdot \left(\sum_{i,j} d_i d_j \langle u_i, u_j\rangle^2 + \frac{1}{k}(\sum_i d_i \|u_i\|^2)^2\right) \\ &\leqslant 32\log^2 k\end{aligned}$$

Therefore $\mathsf{Var}\left[\sum_i d_i f_i^2\right] = \mathbb{E}\left[\sum_{i,j} d_i d_j f_i^2 f_j^2\right] - \left(\mathbb{E}\left[\sum_i d_i f_i^2\right]\right) \leqslant 28\log^2 k$.

□

**Putting It Together**

*Proof of Theorem 1.3.* Now, for each $l \in [k]$, from Lemma 3.2 we get that $\mathbb{E}\left[\sum_i d_i h_l(i)^2\right] = \Theta(\log k)$ and from Lemma 3.4 we get that $\mathsf{Var}\left[\sum_i h_l(i)^2\right] = \Theta(\log^2 k)$. Therefore, from the One-sided Chebyshev inequality (Fact 3.6), we get

$$\mathbb{P}\left[\sum_i d_i h_l(i)^2 \geqslant \frac{\mathbb{E}\left[\sum_i d_i h_l(i)^2\right]}{2}\right] \geqslant \frac{\left(\frac{\mathbb{E}\left[\sum_i d_i h_l(i)^2\right]}{2}\right)^2}{\left(\frac{\mathbb{E}\left[\sum_i d_i h_l(i)^2\right]}{2}\right)^2 + \mathsf{Var}\left[\sum_i h_l(i)^2\right]} \geqslant c'$$

where $c'$ is some absolute constant.

Therefore, with constant probability, for $\Omega(k)$ indices $l \in [k]$, $\sum_i d_i h_l(i)^2 \geqslant \frac{\mathbb{E}\left[\sum_i d_i h_l(i)^2\right]}{2}$. Also, for each $l$, with probability $1 - c'/2$ , $\sum_{i\sim j} |h_l(i)^2 - h_l(j)^2| \leqslant 2/c'\, \mathbb{E}\left[\sum_{i\sim j} |h_l(i)^2 - h_l(j)^2|\right]$. Therefore, with constant probability, for a constant fraction of the indices $l \in [k]$, we have

$$\frac{\sum_{i\sim j} |h_l(i)^2 - h_l(j)^2|}{\sum_i d_i h_l(i)^2} \leqslant \frac{4}{c'} \frac{\mathbb{E}\left[\sum_{i\sim j} |h_l(i)^2 - h_l(j)^2|\right]}{\mathbb{E}\left[\sum_i d_i h_l(i)^2\right]} = \mathcal{O}(\sqrt{\lambda_k \log k})$$

Applying Lemma 3.1 on the vectors with those indices will give $\Omega(k)$ disjoint sets $S_1, \ldots, S_{ck}$ such that $\phi_G(S_i) \leqslant \mathcal{O}(\sqrt{\lambda_k \log k})\ \forall i \in [ck]$.

This completes the proof of Theorem 1.3. □

# 4 Lower bound for $k$ Sparse Cuts

In this Section, we prove Proposition 1.2.

**Proposition 4.1** (Restatement of Proposition 1.2)**.** *For any edge-weighted graph $G = (V, E)$, for any integer $1 \leqslant k \leqslant |V|$, and for any $k$ disjoint subsets $S_1, \ldots, S_k \subset V$*

$$\max_i \phi_G(S_i) \geqslant \frac{\lambda_k}{2}$$

*where $\lambda_1, \ldots, \lambda_{|V|}$ are the eigenvalues of the normalized Laplacian of $G$.*

*Proof.* Let $\alpha$ denote $\max_i \phi_G(S_i)$ Let $T \stackrel{\text{def}}{=} V \backslash (\cup_i S_i)$. Let $G'$ be the graph obtained by shrinking each piece in the partition $\{T, S_i | i \in [k]\}$ of $V$ to a single vertex. We denote the vertex corresponding to $S_i$ by $s_i$ $\forall i$ and the vertex corresponding to $T$ by $t$. Let $\mathcal{L}'$ be the normalized Laplacian matrix corresponding to $G'$. Note that, by construction, the expansion of every set in $G'$ not containing $t$ is at most $\alpha$.

Let $U \stackrel{\text{def}}{=} \{D^{\frac{1}{2}} X_{S_i} | i \in [m]\}$. Here $X_S$ is the incidence vector of the set $S \subset V$. Since all the vectors in $U$ are orthogonal to each other, we have

$$\lambda_k(\mathcal{L}) = \min_{S:rank(S)=k} \max_{x \in S} \frac{x^T \mathcal{L} x}{x^T x} \leqslant \max_{x \in span(U)} \frac{x^T \mathcal{L} x}{x^T x} = \max_{y \in \mathbb{R}^k * \{0\}} \frac{\sum_{i,j} w'_{ij}(y_i - y_j)^2}{\sum_i w'_i y_i^2}$$

For any $x \in \mathbb{R}$, let $x^+ \stackrel{\text{def}}{=} \max\{x, 0\}$ and $x^- \stackrel{\text{def}}{=} \max\{-x, 0\}$. Then it is easily verified that for any $y_i, y_j \in \mathbb{R}$, $(y_i - y_j)^2 \leqslant 2((y_i^+ - y_j^+)^2 + (y_i^- - y_j^-)^2)$. Therefore,

$$\begin{aligned} \textstyle\sum_i \sum_{j>i} w'_{ij}(y_i - y_j)^2 &\leqslant 2(\textstyle\sum_i \sum_{j>i} w'_{ij}(y_i^+ - y_j^+)^2 + \sum_i \sum_{j>i} w'_{ij}(y_j^- - y_i^-)^2) \\ &\leqslant 2(\textstyle\sum_i \sum_{j>i} w'_{ij}|(y_i^+)^2 - (y_j^+)^2| + \sum_i \sum_{j>i} w'_{ij}|(y_j^-)^2 - (y_i^-)^2)| \end{aligned}$$

Without loss of generality, we may assume that $y_1^+ \geqslant y_2^+ \geqslant \ldots \geqslant y_k^+ \geqslant y_t = 0$. Let $T_i = \{s_1, \ldots, s_i\}$ for each $i \in [k]$. Therefore, we have

$$\sum_i \sum_{j>i} w'_{ij} |(y_i^+)^2 - (y_j^+)^2| \leqslant \sum_{i=1}^{k} ((y_i^+)^2 - (y_{i+i}^+)^2) w'(E(T_i, \bar{T}_i)) \leqslant \alpha \sum_{i=1}^{k} ((y_i^+)^2 - (y_{i+i}^+)^2) w'(T_i) = \alpha \sum_i^k w'_i (y_i^+)^2$$

Here we are using the fact that $w'(E(T_i, \bar{T}_i)) \leqslant \alpha w'(T_i)$ which follows from the definition of $\alpha$ and that $w'(T_{i+1}) - w'(T_i) = w'_{i+1}$.

Similiarly, we get that $\sum_i \sum_{j>i} w'_{ij} |(y_j^-)^2 - (y_i^-)^2)| \leqslant \alpha \sum_i w'_i (y_i^-)^2$.

Putting these two inequalities togethor we get that $\sum_{j>i} w'_{ij}(y_i - y_j)^2 \leqslant 2\alpha \sum_i w'_i y_i^2$.

Therefore, $\lambda_k(\mathcal{L}) \leqslant 2 \max_i \phi_G(S_i)$.

□

# 5 $k$-partition

In this Section, we give a constructive proof of Theorem 1.7 The following construction shows that if restrict the $k$-sets $S_1, \ldots, S_k$ to form a partition of $V$, then $\max_i \phi_G(S_i) \gg \Omega(\sqrt{\lambda_k} \mathsf{polylog} k)$.

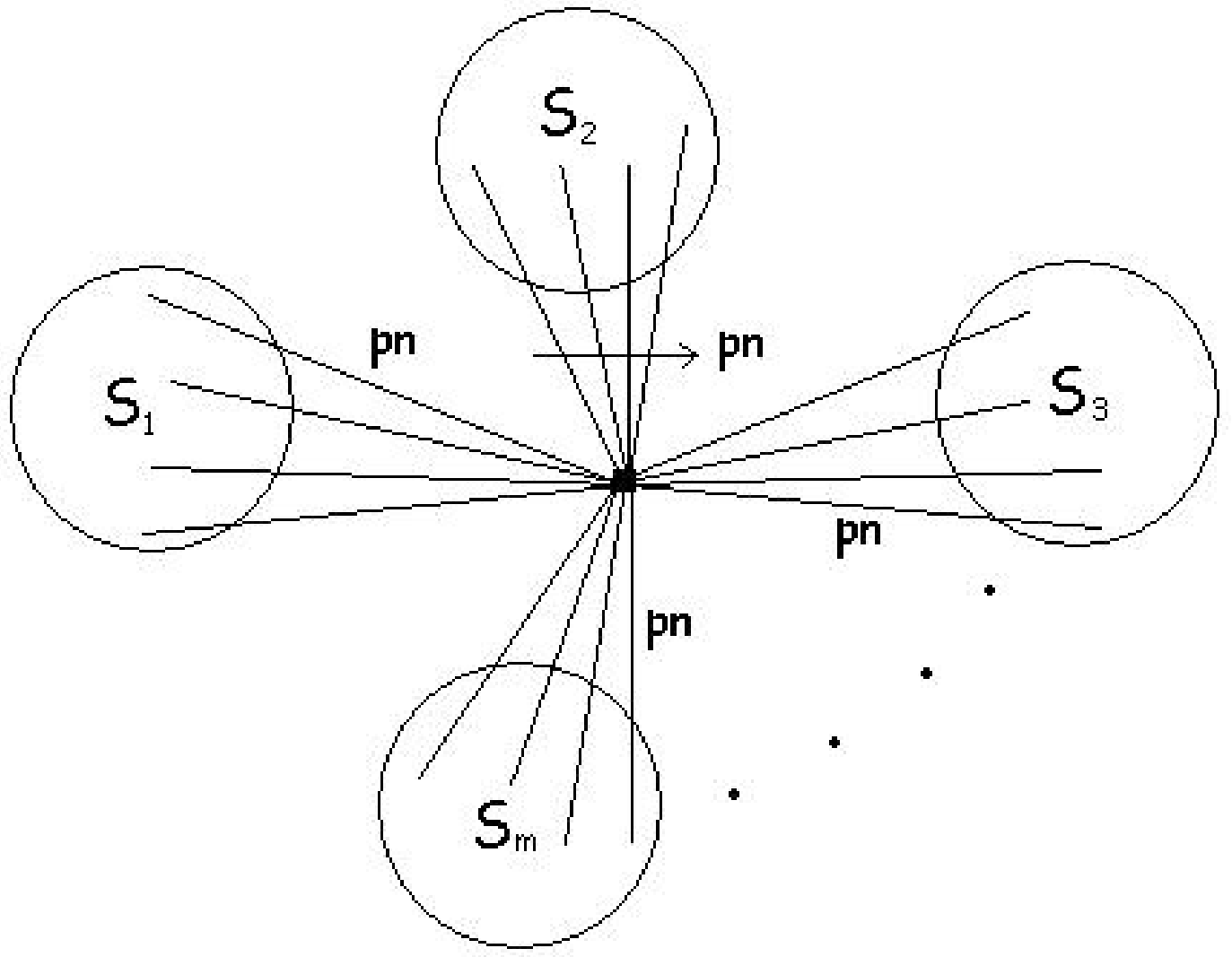

Figure 2: $k$-partition can have sparsity much larger than $\Omega(\sqrt{\lambda_k} \mathsf{polylog} k)$

**Lemma 5.1.** *For the graph $G$ in Figure 2, and for any $k$-partition $S_1, \ldots, S_k$ of its vertex set,*

$$\frac{\max_i \phi_G(S_i)}{\sqrt{\lambda_k}} = \Theta(k^2 \sqrt{\frac{p}{n}})$$

*Proof.* In Figure 2, $\forall i \in [k]$, $S_i$ is a clique of size $(n-1)/k$ (pick $n$ so that $k|(n-1)$). There is an edge from central vertex $v$ to every other vertex of weight $pn$. Here $p$ is some absolute constant.

Let $\mathcal{P}' \stackrel{\text{def}}{=} \{S_1 \cup \{v\}, S_2, S_3, \ldots, S_k\}$. For $n > k^3$, it is easily verified that the optimum $k$-partition is isomorphic to $\mathcal{P}'$.

For $k < o(n^{\frac{1}{3}})$, we have

$$\max_{S_i \in \mathcal{P}'} \phi_G(S_i) = \phi_G(S_1 \cup \{v\}) = \frac{pnk}{\left(\frac{n-1}{k}\right)^2 + pnk} = \Theta\left(\frac{pk^3}{n}\right)$$

Applying Proposition 1.2 to $S_1, \ldots, S_k$, we get that $\lambda_k = \mathcal{O}(pk^2/n)$. Thus we have the Lemma.

□

# A Recursive Algorithms for graph multi-partitioning

The following construction (Figure 3) shows that partition of $V$ obtained using the recursive algorithm of [LRTV11] can give as many as $k(1-o(1))$ sets have expansion $\Omega(1)$ while $\lambda_k \leqslant \mathcal{O}(k^2/n^2)$.

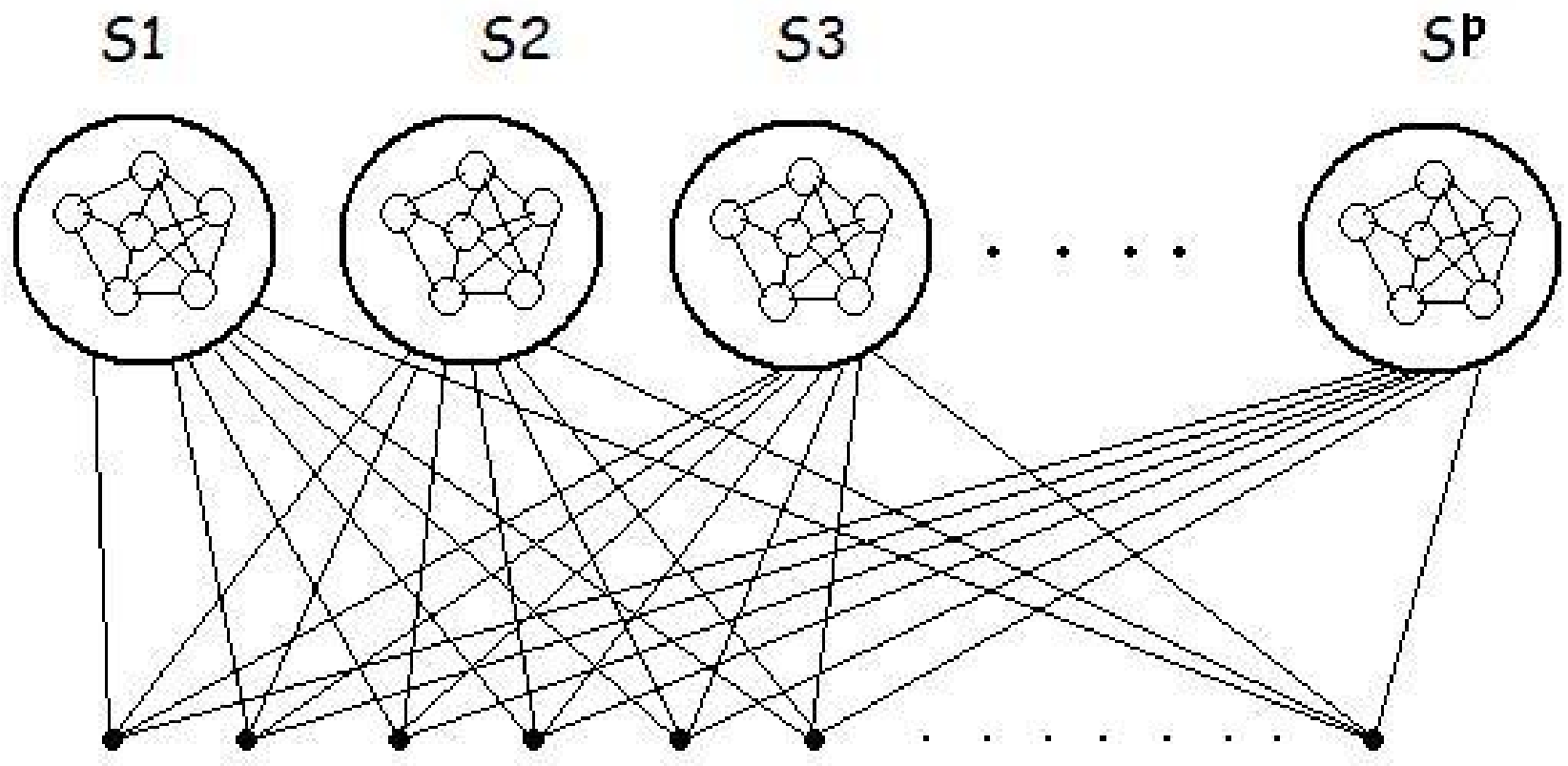

Figure 3: Recursive algorithm can give many sets with very small expansion

In this graph, there are $p \stackrel{\text{def}}{=} k^{\varepsilon}$ sets $S_i$ for $1 \leqslant i \leqslant k^{\varepsilon}$. We will fix the value of $\varepsilon$ later. Each of the $S_i$ has $k^{1-\varepsilon}$ cliques $\{S_{ij} : 1 \leqslant j \leqslant k^{1-\varepsilon}\}$ of size $n/k$ which are sparsely connected to each other. The total weight of the edges from $S_{ij}$ to $S_i \backslash S_{ij}$ is equal to a constant $c$. In addition to this, there are also $k - k^{\varepsilon}$ vertices $v_i : 1 \leqslant i \leqslant k - k^{\varepsilon}$. The weight of edges from $S_i$ to $v_j$ is equal to $k^{-\varepsilon}$.

**Claim A.1.** *1.* $\phi(S_{ij}) \leqslant (c+1)k^2/n^2 \ \forall i,j$

*2.* $\phi(S_i) \leqslant 1/(c+1)\phi(S_{ij}) \ \forall i,j$

*3.* $\lambda_k = \mathcal{O}(m^2/n^2)$

*Proof.* 1.

$$\phi(S_{ij}) = \frac{c + \frac{(m-m^{\varepsilon})m^{-\varepsilon}}{m^{1-\varepsilon}}}{(\frac{n}{m})^2 + c + \frac{(m-m^{\varepsilon})m^{-\varepsilon}}{m^{1-\varepsilon}}} \leqslant \frac{(c+1)m^2}{n^2}$$

2. $w(S_i) = \sum_j w(S_{ij})$, but for each $S_{ij}$ only $1/(c+1)$ fraction of edges incident at $S_{ij}$ are also incident at $S_i$. Therefore, $\phi(S_i) \leqslant 1/(c+1)\phi(S_{ij})$.

3. Follows from (1) and Proposition 1.2.

□

For appropriate values of $\varepsilon$ and $k$, the partition output by the recursive algorithm will be $\{S_i : i \in [k^\varepsilon]\} \cup \{v_i : i \in [k - k^\varepsilon]\}$. Hence, $k(1 - o(1))$ sets have expansion equal to 1.